\documentclass[twocolumn,prc,aps,floatfix]{revtex4}
\usepackage{dcolumn}
\usepackage[dvips]{graphicx}

\begin{document}
\newcommand{\be}{\begin{eqnarray}}
\newcommand{\ee}{\end{eqnarray}}
\title{Effects of the Lorentz invariance violation on Coulomb interactions in nuclei and atoms}
\author{V.V. Flambaum$^{1,2}$},
\author{ M. V. Romalis$^3$}
\affiliation{$^1$School of Physics, University of New South Wales,
Sydney 2052, Australia}
\affiliation{$^2$Helmholtz Institute Mainz, Johannes Gutenberg University, , 55099 Mainz, Germany}
\affiliation{$^3$ Department of Physics, Princeton University, Princeton, New Jersey 08544, USA}
\date{\today}
\begin{abstract}
Anisotropy in the speed of light that has been constrained by Michelson-Morley-type experiments also generates anisotropy in the Coulomb interactions. This anisotropy can  manifest itself as an energy anisotropy in nuclear and atomic experiments. Here the experimental limits on Lorentz violation in $^{21}_{10}$Ne are used to improve the limits on Lorentz symmetry violations in the photon sector, namely the anisotropy of the speed of light and the Coulomb interactions,  by 7 orders of magnitude in comparison with previous experiments: the speed of light is isotropic to a part in $10^{-28}$. 
\end{abstract}
 \maketitle

A special role in the foundation of the theory of relativity was played by the Michelson-Morley experiment searching for the anisotropy in the speed of light. Recent experiments found that the speed of light is isotropic at a level of  $10^{-17} -10^{-21}$ \cite{KosteleckyLIGO,Eisele,Hermann,Hohensee2013c,Pruttivarasin2014}. In Ref.  \cite{Brown2016} it was noted that this limit may be improved to the level of about $10^{-29}$ using the NMR-type experiment \cite{Romalis}. Similar anisotropy in the maximal attainable speed for massive particles has been constrained for nucleons by NMR experiments ( see e.g.  \cite{Romalis,Lamoreaux,Chupp,Prestage,Cs}) and for electrons using optical atomic transitions \cite{Hohensee2013c,Pruttivarasin2014}.  Observation of these or other effects
of the Local Lorentz Invariance Violation (LLIV)  may pave the way to a new, more general  theory  (see e.g.  \cite{KosRus11,BaiCos,KosteleckyPottingPRD1995,Horava,Pospelov,KosLan99,Bluhm,Colladay1997,Colladay1998}).  

Violations of  Lorentz symmetry in the photon sector are parametrized in the Standard Model Extension (SME) \cite{Colladay1998} by the tensor $(k_F)^{\alpha \beta \mu \nu}$. In the presence of Lorentz violation, the Coulomb potential of a point charge becomes anisotropic \cite{BaiCos} :
\begin{equation}\label{eq1}
\Phi({\bf r}) = \frac{q}{r}(1+ (\kappa_{DE})^{ij} n_i n_j/2),
\end{equation}
where   $\kappa_{DE}^{ij}=-2(k_F)^{0 i 0 j}$ are the  tensor components characterizing the  anisotropy in the Coulomb potential and $n_i=x_i/r$ are the unit vectors along the radius-vector in the reference frame in which the components of the $k_F$ tensor are written.

A nucleus that has a finite electric quadrupole moment in the absence of LLIV will exhibit a spatial energy anisotropy due to LLIV caused by the electrostatic interactions of the valence protons with the anisotropic Coulomb potential of the nuclear core.  
The shift of the electrostatic energy due to the LLIV correction in  Eq.~(\ref{eq1}) can be written as  
\begin{equation}\label{E}
\delta E = (\kappa_{DE})^{ij} M_{ij},
\end{equation}
where $M_{ij}$ is the nuclear tensor  which we  calculate here. 
Below we establish a proportionality relation between $M_{ij}$  and the experimental value of the nuclear electric quadrupole moment tensor $Q$, namely $M_{ij}=K Q_{ij}$.   An estimate of the coefficient $K$ allows us to provide estimates of the LLIV shifts for all nuclei and extract the limits on LLIV constants from corresponding experiments. 

We start from a simple analytical estimate of the proportionality coefficient $K$. To obtain the LLIV correction to the electrostatic potential of a finite size charge distribution one has to integrate Eq.~(\ref{eq1}) with a charge density distribution, and the result will not be a simple product of the unperturbed nuclear electrostatic potential $\Phi_0({\bf r})$ and the LLIV factor $(1+ (\kappa_{DE})^{ij} n_i n_j/2)$. However, to clarify the dependence on the parameters of the problem it is instructive to start from a simple estimate assuming  $\Phi({\bf r})  \approx \Phi_0({\bf r})(1+ (\kappa_{DE})^{ij} n_i n_j/2)$.
Below we will also perform an accurate integration of Eq. (\ref{eq1}) with a realistic charge distribution; this gives a  more accurate value of the proportionality factor $K$.
We find that the  LLIV shift of the electrostatic energy $\delta E = (\kappa_{DE})^{ij} M_{ij}$,
may be expressed via the nuclear electric quadrupole moment. Indeed, the quadrupole moment tensor is 
\begin{equation}\label{Qij}
Q_{ij} =\int  ( 3 n_i n_j- \delta _{ij}) \rho({\bf r}) r^2  d^3 r,
\end{equation}
where $\rho({\bf r})$ is the proton number density. The correction to energy produced by the LLIV part of the potential 
$ \Phi_0({\bf r}) (\kappa_{DE})^{ij} n_i n_j/2$ may be presented as $\delta E = (\kappa_{DE})^{ij} M_{ij}$, with 
\begin{equation}\label{Mij}
M_{ij} \approx  \frac{e}{6}\int  ( 3 n_i n_j- \delta _{ij}) \rho({\bf r}) \Phi_0(r)  d^3 r.
\end{equation}
We subtracted $\delta _{ij}$ in the expression above since it does not produce any anisotropic LLIV effect. We see  that the angular factors  in the integrals Eqn.~(\ref{Qij},\ref{Mij}) are the same, we only need to evaluate the radial parts. It is convenient to separate the spherically symmetrical part and the angular-dependent part of the density:
\begin{equation}\label{rho2}
\rho({\bf r}) = \rho_0(r) +  \rho_2({\bf r})
\end{equation}
For a deformed nucleus of a constant density $\rho$ the non-spherical (mainly, quadrupole) part of the density    $\rho_2({\bf r})$ is responsible for the deformation and may be viewed as a correction to the density located near the nuclear surface. Indeed, for a spherical nucleus of a constant density $\rho$  the correction  $\rho_2=  \rho$ outside  the spherical nuclear edge (on the $z$-axis, for the angle $\theta=0$)   makes one axis  longer and $\rho_2=  -\rho$ in the perpendicular direction (for the angle $\theta=\pi/2 $) makes another axis shorter, i.e. $\rho_2({\bf r})$ transforms the spherical nucleus into the spheroidal one.  The electrostatic potential near the nuclear surface is equal to  $\Phi_0({\bf r}) \approx Ze/R$ and is continuous across the surface. So, for small quadrupolar deformations 
 we obtain from Eqn. (\ref{Qij},\ref{Mij}) the relation between  $M=M_{zz}$ and $Q=Q_{zz}$: 
\begin{equation}\label{MQs}
M_{estimate} \sim \frac{1}{6} \frac{(Z-1)e^2 Q}{R^3}
\end{equation}
 Note that we replaced the nuclear charge factor $Z$ by $Z-1$ to account for the charge quantization. Indeed, to exclude the interaction of a proton with itself the result must vanish for $Z=1$. Transformation from the frozen body frame to the laboratory reference frame does not change the ratio $M/Q$.

We have also performed  numerical calculations of the correction to the energy due to the LLIV part of the electrostatic potential Eq. (\ref{eq1}) for the interaction of the point charges using the accurate numerical integration
\begin{equation}
\delta E = \frac{1}{2}\int \Phi({\bf r_1} -  {\bf r_2})\rho({\bf r_1})\rho({\bf r_2}) d^3 {\bf r_1} d^3{\bf r_2}. 
\end{equation}
 
For the density $\rho({\bf r})$ we assume the model of the spheroidal nucleus with a constant charge density and the nuclear radius given by $R=R_0 (1+\beta Y_{20}(\theta,\phi)))$.  We have found that  the numerical results in this model for nuclei with different $R$, $Z$ and  $Q$ may be approximated by the analytical  formula
\begin{equation}\label{MQnumerical }
M  =0.06 \frac{(Z-1)e^2 Q}{R^3}=0.055  \frac{Z-1}{A}\frac{Q}{{\rm fm}^2}{\rm MeV}.
\end{equation}  
We assume the nuclear radius $R=1.15 A^{1/3}$ fm,  where $A$ is the number of nucleons. The ratio $M/Q$ changes by about $\pm10\%$ as $\beta$ changes from 0 to 0.4, as shown in Fig. 1,  so the result can be applied to all nuclei.  The experimental values of the nuclear quadrupole moments $Q$ may be found in the tables   \cite{Tables}.  For the  $^{21}_{10}$Ne nucleus $Q=$10.3 fm$^2$, and we obtain   $M=0.25 $ MeV.

\begin{figure}
	\includegraphics[width=3.2in]{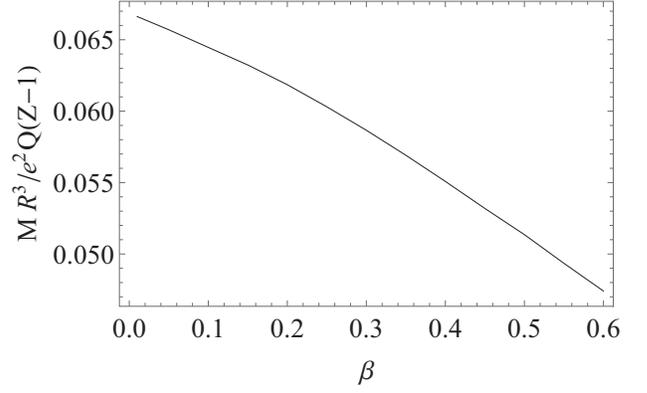}
	\caption{Dependence of the ratio $M/Q$ on the nuclear quadrupole deformation size $\beta$.}\label{Fig1}
 \end{figure}

Now we can extract the limits on the  coefficients $(\kappa_{DE})^{ij}$ responsible for the  asymmetry in the speed of light from the measurements of LLIV in the  $^{21}_{10}$Ne nucleus \cite{Romalis}. We are interested in two contributions,     
 \begin{equation}\label{Ekc}
\delta E =(\kappa_{DE})^{ij} M_{ij} - c_{ij}P_{ij} \,\, .
\end{equation}
The coefficients $c_{ij}$ characterize anisotropy in the maximal attainable speed for massive particles
 or the anisotropy in the kinetic energy term, which in the non-relativistic limit may be presented as  $-c_{ij}P_{ij}$, as derived in \cite{Clockcom}. Here the tensor $P_{ij}=\left \langle 3 p_i p_j- p^2 \delta_{ij} \right \rangle/6m$,  $p$ and $m$ are the particle momentum and mass.  In  Ref. \cite{Romalis}  the limits  on $c_{ij}$ for neutrons were obtained assuming $(\kappa_{DE})^{ij}=0$:
\begin{eqnarray}
c^{X}_n&=&c^{YZ}_n+c^{ZY}_n=(4.8 \pm 4.4) \cdot 10^{-29},\nonumber\\ 
c^{Y}_n&=&c^{XZ}_n+c^{ZX}_n=-(2.8 \pm 3.4) \cdot 10^{-29},\nonumber\\ 
c^{Z}_n&=&c^{XY}_n+c^{YX}_n=-(1.2 \pm 1.4) \cdot 10^{-29},\nonumber\\
c^{-}_n&=&c^{XX}_n-c^{YY}_n=(1.4 \pm 1.7) \cdot 10^{-29}.\\
\end{eqnarray}

 The limits on the $c_{ij}$ above have been obtained assuming the Schmidt (single valence neutron) value $P^{n}=P^{n}_{zz}=-0.66$ MeV  \cite{Romalis}. More accurate  calculations of $P$ \cite{Flambaum2016,Brown2016} have shown that both neutrons and protons contribute to   $c_{ij}$. According to Ref. \cite{Flambaum2016}  $P^{p}=0.54$ MeV, $P^{n}=0.57$ MeV  and the experimental limits on  $c_{ij}$
  actually contain the linear combination $c= - (0.8 c^{p} + 0.9 c^{n})$. In Ref. \cite{Brown2016} the recommended values for the matrix elements are  $P^{p}=0.47 $ MeV and $P^{n}= 0.7$ MeV, so the results of the two approaches are quite consistent.   We need to add  the photon contribution $(\kappa_{DE})^{ij}$ calculated in the present work with  $M=0.25 $ MeV. This gives the linear combination of the coefficients that are constrained by the experimental measurements in $^{21}$Ne presented above: 
  
  \begin{eqnarray}
  c^{X}_n+0.9 c^{X}_p-0.4 \kappa^{X}_{DE}&=&-(5.3\pm 4.9) \cdot 10^{-29},\nonumber \\
  c^{Y}_n+0.9 c^{Y}_p-0.4 \kappa^{Y}_{DE}&=&(3.1\pm 3.8) \cdot 10^{-29}, \nonumber\\
  c^{Z}_n+0.9 c^{Z}_p-0.4 \kappa^{Z}_{DE}&=&(1.3\pm 1.6) \cdot 10^{-29}, \nonumber\\
  c^{-}_n+0.9 c^{-}_p-0.4 \kappa^{-}_{DE}&=&-(1.6\pm 1.9) \cdot 10^{-29}.\label{NeC} 
  \end{eqnarray}

It is instructive to compare these calculations to the case of Coulomb interaction in atoms. 
For example, for  an electron in a one-electron atom, one can write the main LLIV energy contributions as 
\begin{equation}
\delta E  =- \frac{Ze^2}{r}\frac{(\kappa_{DE})^{ij} n_i n_j}{2}-c_{ij} p_i p_j/m
\end{equation}
The two terms can be related to each other using a generalization of the virial theorem. In particular for an atomic eigenstate
\begin{equation}
\frac{d \left \langle p_i r_j \right \rangle }{dt}=0=\frac{i}{\hbar}\left\langle [H,p_i r_j]\right \rangle
\end{equation}
where $H=-Ze^2/r +p_i p_i/2m$. It follows that 
\begin{equation}
\left\langle\frac{p_i p_j}{m}\right\rangle= \left\langle\frac{Z e^2 n_i n_j}{r}\right\rangle \label{Virial}
\end{equation}
The virial theorem is valid in many-electron atoms too.  We should include double sum over interaction energies of all particles into the Hamiltonian and make corresponding changes in the equations above. Therefore, for any atom with the electron energy dominated by the non-relativistic Hamiltonian, 
the two LLIV terms always enter as a combination
$(\kappa_{DE})^{ij}/2+c_{i,j}$. 

A similar result can be derived using a coordinate transformation $x^{\mu} \rightarrow x^{\mu}-(k_F)^{\alpha\mu}_{\quad \alpha\nu} x^{\nu}/2$  which removes Lorentz violation in the photon sector and modifies the electron LLIV coefficients $c_{\mu\nu}\rightarrow c_{\mu\nu}+(k_F)^{\alpha}_{\;\;\mu\alpha\nu}/2$, as discussed in \cite{KosteleckyTasson}. Using Riemann tensor symmetries of the $(k_F)^{\alpha}_{\;\;\mu\alpha\nu}$ tensor one can show that  $(k_F)^{\alpha}_{\;\;j\alpha k}=(\tilde{\kappa}_{e^-})^{jk}$ for the traceless components of the tensor. Furthermore, by imposing stringent constraints on birefringence components of the photon LLIV tensor $k_{e^+}$ at a level of $10^{-38}$  \cite{Mewes}, one obtains $(\kappa_{DE})^{ij}=(\tilde{\kappa}_{e^-})^{ij}$ for the traceless components. Hence in an atom held together by Coulomb interactions only one linear combination of the $(\tilde{\kappa}_{e^-})^{ij}$ and $c_{ij}$ tensors can be measured. This relationship based on a coordinate transformation was used to relate the photon sector coefficients to the electron $c_{ij}$ coefficients in \cite{Pruttivarasin2014}.

The situation is different in the nucleus, where the relationship (\ref{Virial}) does not hold since the total potential energy is dominated by strong interactions. Note that the linear combination  in Eqs.~(\ref{NeC}) has the opposite sign between the nucleon and photon coefficients because the Coulomb potential energy is positive in the nucleus, unlike an atom. Our calculations implicitly assume that there is no LLIV in the strong interaction.
 Existing limits on Lorentz violation in strong interactions come from ultra-high energy particles of astrophysical origin, in particular  limits  on the difference between the maximum attainable velocity for quarks and gluons, which are on the order of $10^{-21}$ to $10^{-24}$  \cite{Moore,Carone}. In general, as discussed in \cite{KosteleckyTasson}, Lorentz violation can be only defined in the differences in the properties of two types of particles. It is convenient therefore to use the freedom of coordinate transformation to remove  LLIV for strong interactions. In this coordinate system, the relationships $(\ref{NeC})$ can be interpreted as providing a constraint on the sum of Lorentz violation for nucleons and photons.   

Modern analogues of the Michelson-Morley experiments \cite{KosteleckyLIGO,Eisele,Hermann} measure the anisotropy in the speed of light relative to the length of a material cavity. Lorentz violation due to changes in the cavity length have been considered in \cite{Holger}. Such analyses use a coordinate system where the centers of mass of the nuclei in the material lattice are at rest on average and have a kinetic energy on the order of the Debye temperature, much less than 1 eV. Furthermore, the internal motion of the nucleons inside each nucleus has no preferential direction since the nuclei are not spin-polarized as in the $^{21}$Ne experiment \cite{Romalis} and the expectation value of the $P_{ij}$ tensor is nearly zero \cite{NQR}. This largely suppresses the Lorentz-violating effects from the anisotropy of the kinetic energy of protons and neutrons. Therefore Michelson-Morley style experiments place constraints primarily on the combination of the electron and the photon coefficients. To compare our limits to previous measurements of LLIV in the photon sector it is natural therefore to assume  $c^{p}_{ij}=c^{n}_{ij}=c^{e}_{ij}=0$. With these assumptions we obtain:
\begin{eqnarray}
\tilde{\kappa}_{e^-}^X&=&\tilde{\kappa}_{e^-}^{YZ}+\tilde{\kappa}_{e^-}^{ZY}=-(13 \pm 12) \cdot 10^{-29},\nonumber\\ 
\tilde{\kappa}_{e^-}^Y&=&\tilde{\kappa}_{e^-}^{XZ}+\tilde{\kappa}_{e^-}^{ZX}=(7 \pm 9) \cdot 10^{-29},\nonumber\\ 
\tilde{\kappa}_{e^-}^Z&=&\tilde{\kappa}_{e^-}^{XY}+\tilde{\kappa}_{e^-}^{YX}=-(3.2 \pm 3.7) \cdot 10^{-29},\nonumber\\
\tilde{\kappa}_{e^-}^-&=&\tilde{\kappa}_{e^-}^{XX}-\tilde{\kappa}_{e^-}^{YY}=-(3.7 \pm 4.5)  \cdot 10^{-29},
\label{NeK}
 \end{eqnarray}
These limits  are  7 - 11 orders of magnitude better than that 
obtained in the previous  laboratory measurements \cite{KosteleckyLIGO,Eisele,Hermann,Hohensee2013c,Pruttivarasin2014}.

To summarize, in this work we performed calculations which  give  a new interpretation of existing and future experiments searching for the anisotropy in the speed of light and the anisotropy in the maximal attainable speed for massive particles. 
New interpretation of the experiment with $^{21}$Ne \cite{Romalis} has already allowed us to improve the limits on some LLIV interaction parameters for photons describing asymmetry in the speed of light (studied in the Michelson-Morley experiment) and anisotropy in the Coulomb interaction  by 7 orders of magnitude.

We are grateful  to A. Kostelecky and M. Hohensee for valuable discussions. This work is  supported in part by the Australian Research Council and the Gutenberg Fellowship (J. Gutenberg University, Mainz, Germany).
and the National Science Foundation, USA.


\begin{thebibliography}{99}
\bibitem{NQR} In single crystals nuclei with an electric quadrupole moment will be polarized at a level of $10^{-7}$ at room temperature.
\end{thebibliography}
\end{document}